\documentclass[12pt]{article}
\makeatletter \@addtoreset{equation}{section} \makeatother

\newcommand{\noi}{\vspace{12pt}\noindent}
\newcommand{\beq}{\begin{equation}}
\newcommand{\eeq}{\end{equation}}
\newcommand{\bea}{\begin{eqnarray}}
\newcommand{\eea}{\end{eqnarray}}
\newcommand{\e}[1]{{(\ref{#1})}}
\newcommand{\eq}[1]{{eq.\ (\ref{#1})}}
\newcommand{\es}[2]{{(\ref{#1}) and (\ref{#2})}}
\newcommand{\eqs}[2]{{eqs.\ (\ref{#1}) and (\ref{#2})}}
\newcommand{\Ref}[1]{{Ref.~\cite{#1}}}

\newcommand{\equi}[1]{\stackrel{{#1}}{=}}

\newcommand{\ie}{{${ i.e., \ }$}}
\newcommand{\eg}{{${ e.g., \ }$}}

\newcommand{\cf}{{cf.\ }}

\newcommand{\Hf}{{1 \over 2}}
\newcommand{\ih}{{\scriptstyle{i \over \hbar}}}

\newcommand{\gh}{{\rm gh}}
\newcommand{\tr}{{\rm tr}}

\newcommand{\vol}{{\rm vol}}
\newcommand{\for}{{\rm for}}
\newcommand{\diag}{{\rm diag}}
\newcommand{\brst}{{\rm\bf s}}
\newcommand{\EOM}{E}
\newcommand{\cA}{{\cal A}}
\newcommand{\nc}{{non--commutative }}
\newcommand{\BV}{{Batalin--Vilkovisky }}
\newcommand{\eps}{\varepsilon^{}}
\renewcommand{\tilde}{\widetilde}
\renewcommand{\bar}{\overline}
\renewcommand{\Re}{{\rm Re}}
\renewcommand{\Im}{{\rm Im}}

\newcommand{\twobytwo}[4]{\left(\begin{array}{ccc}{#1}&{#2} \cr
                                  {#3} & {#4} \end{array} \right)}
\newcommand{\papar}[1]{{
 {\stackrel{\raise.2ex\hbox{$\leftarrow$}}{\partial^{r}}   }
\over {   \partial {#1}}  }}
\newcommand{\papal}[1]{{
 {\stackrel{\lower.3ex \hbox{$\rightarrow$}}{\partial^{\ell}}   }
 \over {   \partial {#1}}  }}

\newcommand{\proofbox}{\begin{flushright}
${\,\lower0.9pt\vbox{\hrule \hbox{\vrule
height 0.2 cm \hskip 0.2 cm \vrule height 0.2 cm}\hrule}\,}$
\end{flushright}}

\begin{document}
\thispagestyle{empty}
\title{\Large{\bf On Batalin--Vilkovisky Formalism\\
of Non--Commutative Field Theories}}
\author{{\sc Klaus~Bering}$^{a}$ and {\sc Harald~Grosse}$^{b}$ \\~\\
$^{a}$Institute for Theoretical Physics \& Astrophysics\\
Masaryk University\\Kotl\'a\v{r}sk\'a 2\\CZ--611 37 Brno\\Czech Republic\\~\\
$^{b}$Department of Physics\\University of Vienna\\
Boltzmanngasse 5\\A--1090 Vienna\\Austria}
\maketitle
\vfill
\begin{abstract}
We apply the BV formalism to non--commutative field theories, introduce BRST 
symmetry, and gauge--fix the models. Interestingly, we find that treating the
full gauge symmetry in non--commutative models can lead to reducible gauge 
algebras. As one example we apply the formalism to the Connes--Lott 
two--point model. {}Finally, we offer a derivation of a superversion of the 
Harish--Chandra--Itzykson--Zuber integral.
\end{abstract}
\vfill
\begin{quote}
PACS number(s): 02.40.Gh; 03.65.Ca; 11.10.-z; 11.10.Gh; 11.10.Nx; 11.15.-q. 
\hrule width 5.cm \vskip 2.mm \noindent
$^{a}${\small E--mail:~{\tt bering@physics.muni.cz}} \hspace{10mm}
$^{b}${\small E--mail:~{\tt harald.grosse@univie.ac.at}} \\
\end{quote}

\section{Introduction}

\noi 
Developments around field theory models defined over \nc spaces are impressive.
The formulation of various kinds of models is possible and was especially
boosted after the paper \cite{Seiberg:1999vs}. The main hope to cure the
diseases of quantum field theory was, however, only partially fulfilled. The
canonical deformation leads to the IR/UV mixing. 

\noi
{}For a \nc scalar field theory a detailed rigorous treatment of R.~Wulkenhaar
and one of the authors (H.G.) led to the identification of four
relevant/marginal operators and a renormalizability proof \cite{Grosse:2004yu}.
The resulting model has the nice feature that the beta--function of the
coupling constant vanishes to all orders of perturbation theory, which may
lead to a constructive procedure \cite{Grosse:2004by,Disertori:2006nq}. {}For
a beautiful review of this subject with many references, see 
\cite{Rivasseau:2007ab}.

\noi 
Non--commutative gauge models have been treated first by expanding in the
deformation parameter and using the Seiberg--Witten map
\cite{Jurco:2001rq,Bichl:2001cq}. The treatments without expansions are
extensive, but the question of renormalizability of these gauge models has
been answered only partially, see, \eg the proposals
\cite{de Goursac:2007gq,Grosse:2007dm} resulting from a heat kernel expansion.
In addition, a Becchi--Rouet--Stora--Tyutin (BRST) approach was developed for
a specific model
\cite{Blaschke:2007vc} such that all propagators have nice decay properties
resulting from a coupling to an oscillator term. Loop calculations in this
specific model indicate improvements over elder models, but no conclusion for
renormalization up to all orders has been possible. There has also been a
recent attempt of using a different type of non--local counter--term in
\cite{Vilar:2009er}. In this way it is possible to yield what is called
localization, see \cite{Blaschke:2009hp} for a recent treatment, but even this
approach is still not conclusive.

\noi 
Many of these \nc systems are matrix models with a cutoff given by the matrix
size. Removing the cutoff leads to infinite gauge volume for gauge models.
Therefore it is necessary to gauge--fix before taking the infinite matrix
limit. This led us to study gauge models on matrix algebras including
gauge--fixing, which is the main topic of this letter. We find that the \BV
(BV) formalism \cite{Batalin:1981jr,Batalin:1984jr} is here a useful (and in
many instances a necessary) tool. 

\noi 
The letter is organized as follows. In Section~\ref{secderham}, we discuss a
construction of a \nc de Rham differential that works both for Heisenberg
algebra type and Lie algebra type of non--commutativity. In
Sections~\ref{sectargetspace}--\ref{secgf} we formulate \nc gauge theories
in the BRST and the BV formalism. The gauge algebra can be reducible
\cite{Batalin:1984jr,Batalin:2009fe}, but gauge--fixing is still possible. In
Section~\ref{secconneslott} we apply the stage--one reducible BV formalism to
the Connes--Lott \nc model \cite{Connes:1990qp}, which has built in the Higgs
effect. {}Finally, in Appendix~\ref{seciz} we give a localization argument for
the superversion of the Harish--Chandra--Itzykson--Zuber integral.

\noi
We expect that the \BV formalism can be applied to many other models of \nc 
quantum field theory, particularly when analyzing renormalizability, and we 
shall consider more applications in the future.

\noi 
{\sc General Remarks About Notation}: Adjectives from super--mathematics such
as ``gra\-ded'', ``super'', etc., are implicitly implied. The commutator
$[f,g]$ of two \nc forms $f$ and $g$, of Grassmann--parity $\eps_{f}$,
$\eps_{g}$ and of form--degree $p^{}_{f}$, $p^{}_{g}$, is defined as 
\beq 
[f,g] = f g -(-1)^{\eps_{f}\eps_{g}+p^{}_{f}p^{}_{g}} g f .
\label{supercommutator} 
\eeq 

\noi
There is a tradition in quantum mechanical textbooks to put a hat ``$\wedge$'' 
on top of a \nc operator $\hat{f}$, to distinguish it from  its commutative 
symbol $f$, which is just a function. However, we shall not write hats 
``$\wedge$'' to avoid clutter. The commutative symbol will only appear in 
eqs.\ \e{moyal}, \es{fint}{ffct} below.

\noi
{}Finally, we should mention that we do often not discuss reality/Hermiticity
conditions explicitly. Since we will often have no explicit factors of the
imaginary unit $\sqrt{-1}$ in our formulas, we should warn that the variables
are sometimes implicitly assumed to be imaginary/anti--Hermitian rather than
real/Hermitian.

\section{Non--Commutative de Rham Differential}
\label{secderham}

\noi
Let there be given an associative algebra $\cA$ with algebra generators
$x^{}_{\mu}$, $\mu\!\in\!I$, and a unit ${\bf 1}$. It is assumed that the set
$\{{\bf 1}\}\cup\{x^{}_{\mu}|\mu \in I\}$ consists of linearly independent
elements. Physically, we can think of the algebra $\cA$ as a \nc world
volume with \nc coordinates $x^{}_{\mu}$. We will often realize the 
$x^{}_{\mu}$ coordinates as matrices $(x^{}_{\mu})^{a}{}^{}_{b}$, where the 
matrix index ``$a$'' carries Grassmann--parity $\eps_{a}$, so that the matrix 
entry $(x^{}_{\mu})^{a}{}^{}_{b}$ has Grassmann--parity
\beq
\eps((x^{}_{\mu})^{a}{}^{}_{b})=\eps_{\mu}+\eps_{a}+\eps_{b} .
\eeq
We will also assume that there exists a cyclic trace operation ``$\tr$'' for
the algebra $\cA$. The trace operation ``$\tr$'' may be thought of as an
integration over the \nc world volume. In a matrix realization, the trace
``$\tr$'' is the supertrace,
\beq
\tr(x^{}_{\mu})=(-1)^{\eps_{a}(\eps_{\mu}+1)} (x^{}_{\mu})^{a}{}^{}_{a} .
\eeq
We next assume that the commutator $[x^{}_{\mu}, x^{}_{\nu}]$ of two coordinates
$x^{}_{\mu}$ and $x^{}_{\nu}$ is a linear combination of
$\{{\bf 1}\}\cup\{x^{}_{\mu}|\mu \in I\}$, \ie that there exists antisymmetric
structure constants
\bea
\theta^{}_{\mu\nu}&=&-(-1)^{\eps_{\mu}\eps_{\nu}}\theta^{}_{\nu\mu} ,
\label{theta} \\
f^{}_{\mu\nu}{}^{\lambda}
&=&-(-1)^{\eps_{\mu}\eps_{\nu}}f^{}_{\nu\mu}{}^{\lambda} , 
\label{f1}
\eea such that
\beq [x^{}_{\mu}, x^{}_{\nu}]
= \theta^{}_{\mu\nu} {\bf 1}+ f^{}_{\mu\nu}{}^{\lambda}x^{}_{\lambda} .
 \label{xxfx1}
\eeq
This will cover two main applications: the Heisenberg algebra, \ie the
constant case with $f^{}_{\mu\nu}{}^{\lambda}\!=\!0$; and the Lie algebra, \ie
the linear case with $\theta^{}_{\mu\nu}\!=\!0$. The Jacobi identity for 
commutator $[\cdot,\cdot]$ and the linear independence imply that
\bea
\sum_{{\rm cycl.}~\mu,\nu,\lambda}(-1)^{\eps_{\mu}\eps_{\lambda}}
f^{}_{\mu\nu}{}^{\kappa}\theta^{}_{\kappa\lambda} &=& 0 , \label{jacid1} \\
\sum_{{\rm cycl.}~\mu,\nu,\lambda}(-1)^{\eps_{\mu}\eps_{\lambda}}
f^{}_{\mu\nu}{}^{\kappa}f^{}_{\kappa\lambda}{}^{\rho} &=& 0 . \label{jacid2}
\eea
One next defines a (not necessarily nilpotent) Bosonic de Rham one--form
\beq
\Omega = c^{\mu} x^{}_{\mu}
+ \Hf c^{\nu} c^{\mu} f^{}_{\mu\nu}{}^{\lambda}b^{}_{\lambda} .
\eeq
Here the $c^{\mu}$'s and the $b^{}_{\mu}$'s are bases for one--forms and
minus--one--forms(=vector fields), respectively.
\beq
[ b^{}_{\mu},c^{\nu} ] = {\delta}_{\mu}^{\nu}, \label{bccom}
\eeq
and all other commutators vanish. The form degree ``$p$'' can be thought of as
a world volume ghost degree, and in this sense, the $c^{\mu}$'s and the
$b^{}_{\mu}$'s are world volume ghosts and ghost momenta. (This should not be
confused with the actual ghost number ``$\gh$'', which lives in a target
space.) 

\noi
The components $\Omega^{}_{\mu}$ of the de Rham one--form
$\Omega\!=\!c^{\mu}\Omega^{}_{\mu}$ is
\beq
\Omega^{}_{\mu}
= x^{}_{\mu} + \Hf c^{\nu} f^{}_{\nu\mu}{}^{\lambda}b^{}_{\lambda} .
\eeq
The square
\beq
\Omega^{2}=\Hf[\Omega,\Omega]
= -\Hf c^{\nu}c^{\mu}\theta^{}_{\mu\nu}
= \Hf c^{\mu}\theta^{}_{\mu\nu}c^{\nu}(-1)^{\eps_{\nu}}
\eeq
of the de Rham one--form $\Omega$ is a (not necessarily vanishing) two--form.
The \nc exterior de Rham differential $d$ is now implemented as
\beq
d := [\Omega, \cdot\ ] .
\eeq
The square
\beq d^2 = [\Omega, [\Omega,\cdot\ ]] = [\Omega^{2}, \cdot\ ]
\eeq
of the de Rham differential ``$d$'' vanishes on elements
$F\!=\!F(x,c)\!\in\!\Omega^{\bullet}(\cA)$
that do not depend on the minus--one--forms $b^{}_{\mu}$.

\begin{table}
\caption{Parities, degrees and ghost numbers of various objects.}
\label{gradtable}
\begin{center}
\begin{tabular}[t]{|l|c||c|c|c|} \hline
&&Grass-&World&Target \\
&&mann&volume&space \\
&&parity&form&ghost \\
&&&degree&number \\ \hline
&$\downarrow$ \ Symbol \ $\to$&$\eps$&$p$&$\gh$\\ \hline\hline
World volume coordinate&$x^{}_{\mu}$&$\eps_{\mu}$&$0$&$0$\\ \hline
World volume one--form&$c^{\mu}$&$\eps_{\mu}$&$1$&$0$\\ \hline
World volume minus--one--form&$b^{}_{\mu}$&$\eps_{\mu}$&$-1$&$0$\\
\hline\hline
De Rham one--form&$\Omega=c^{\mu}\Omega^{}_{\mu}$&$0$&$1$&$0$\\ \hline
De Rham differential &$d=[\Omega,\cdot]$&$0$&$1$&$0$\\  \hline\hline
General target space field&$\Phi^{\alpha}$
&$\eps_{\alpha}$&$0$&$\gh_{\alpha}$\\ \hline
Target space coordinate&$X^{}_{\mu}\!=\!x^{}_{\mu}\!+\!A^{}_{\mu}$
&$\eps_{\mu}$&$0$&$0$\\ \hline
Gauge parameter&$\Xi$&$0$&$0$&$0$\\ \hline
Target space ghost&$C$&$1$&$0$&$1$\\ \hline
Target space antighost\rule{0ex}{2.5ex}&$\bar{C}$&$1$&$0$&$-1$\\ \hline
Lagrange multiplier&$\Pi$&$0$&$0$&$0$\\ \hline
Gauge condition&$\chi$&$0$&$0$&$0$\\ \hline\hline
Ghost--for--ghost &$\eta$&$0$&$0$&$2$\\ \hline
Antighost--for--ghost &$\bar{\eta}$&$0$&$0$&$-2$\\ \hline
Lagr.--mult.--for--ghost &$\bar{\pi}$&$1$&$0$&$-1$\\ \hline
Extra ghost &$\tilde{\eta}$&$0$&$0$&$0$\\ \hline
Extra Lagrange multiplier
&$\tilde{\pi}$&$1$&$0$&$1$\\ \hline\hline
General target space antifield&$\Phi^{*}_{\alpha}$
&$\eps_{\alpha}\!+\!1$&$0$&$-1\!-\!\gh_{\alpha}$ \\ \hline
Coordinate antifield&$X^{\mu*}$&$\eps_{\mu}\!+\!1$&$0$&$-1$\\ \hline
Ghost antifield&$C^{*}$&$0$&$0$&$-2$\\ \hline
Antighost antifield\rule{0ex}{2.5ex}&$\bar{C}^{*}$&$0$&$0$&$0$\\ \hline
Lagrange multiplier antifield&$\Pi^{*}$&$1$&$0$&$-1$\\ \hline\hline
Ghost--for--ghost antifield&$\eta^{*}$&$1$&$0$&$-3$\\ \hline
Antighost--for--ghost antifield&$\bar{\eta}^{*}$&$1$&$0$&$1$\\ \hline
Lagr.--mult.--for--ghost antifield&$\bar{\pi}^{*}$&$0$&$0$&$0$\\ \hline
Extra ghost antifield&$\tilde{\eta}^{*}$&$1$&$0$&$-1$\\ \hline
Extra Lagr.mult.\ antifield
&$\tilde{\pi}^{*}$&$0$&$0$&$-2$\\ \hline\hline
Classical BRST operator&$\brst=(S,\cdot)$&$1$&$0$&$1$\\ \hline
Odd Laplacian&$\Delta$&$1$&$0$&$1$\\ \hline
Gauge--fermion&$\Psi$&$1$&$0$&$-1$\\ \hline
\end{tabular}
\end{center}
\end{table}

\section{Non--Commutative Gauge Field Models}
\label{sectargetspace}

\noi 
{}For these models it is possible to introduce a one--form valued covariant
derivative 
\beq 
\nabla = \Omega + A = c^{\mu}\nabla^{}_{\mu}, 
\eeq 
where the one--form $A\!=\!c^{\mu}A^{}_{\mu}$ is a
gauge potential. One usually assumes that the gauge field
components $A^{}_{\mu}\!=\!A^{}_{\mu}(x)$ do not depend on the $c$'s and
$b$'s. The components $\nabla^{}_{\mu}$ of the covariant derivative
$\nabla$ are 
\beq 
\nabla^{}_{\mu} = \Omega^{}_{\mu} + A^{}_{\mu} = X^{}_{\mu}
+ \Hf c^{\nu} f^{}_{\nu\mu}{}^{\lambda}b^{}_{\lambda} , 
\eeq 
where 
\beq X^{}_{\mu} := x^{}_{\mu} + A^{}_{\mu} 
\eeq 
are the covariant coordinates. One can think of $X^{}_{\mu}=X^{}_{\mu}(x)$ as 
coordinates on a target space. The field strength $F$ and the curvature $R$ 
are defined as
\bea 
{}F &:=& (dA)+A^{2}=-\Hf c^{\nu}c^{\mu}F^{}_{\mu\nu}
= \Hf c^{\mu}F^{}_{\mu\nu}c^{\nu}(-1)^{\eps_{\nu}} , \\
R&:=&\nabla^2=\Hf[\nabla,\nabla]=\Omega^{2}+F
=-\Hf c^{\nu}c^{\mu}R^{}_{\mu\nu}
= \Hf c^{\mu}R^{}_{\mu\nu}c^{\nu}(-1)^{\eps_{\nu}},
\eea
respectively. Their components $F^{}_{\mu\nu}$ and $R^{}_{\mu\nu}$ do not depend
on the $c$'s and $b$'s.
\bea
{}F^{}_{\mu\nu}&=&[x^{}_{\mu},A^{}_{\nu}]+[A^{}_{\mu},x^{}_{\nu}]
+[A^{}_{\mu},A^{}_{\nu}]-f^{}_{\mu\nu}{}^{\lambda}A^{}_{\lambda}, \\
R^{}_{\mu\nu}&=&F^{}_{\mu\nu}+\theta^{}_{\mu\nu}
=[X^{}_{\mu},X^{}_{\nu}]-f^{}_{\mu\nu}{}^{\lambda}X^{}_{\lambda} .
\eea
The typical starting action $S^{}_{0}$ is of the form
$S^{}_{0}\!=\!\tr L^{}_{0}(X)$, where $L^{}_{0}\!=\!L^{}_{0}(X)$ is a polynomial
in the $X^{}_{\mu}$'s. The covariant coordinates $X^{}_{\mu}$ transform as
$X^{}_{\mu}\to X_{\mu}^{g}\!=\!g^{-1}X^{}_{\mu}g$ under gauge transformations
$g=e^{\Xi}$. Therefore the infinitesimal gauge transformations takes the form
\beq
\delta X^{}_{\mu} = [X^{}_{\mu}, \Xi] = -[\Xi, X^{}_{\mu}] , \label{gaugesym1}
\eeq
where $\Xi\!\in\!\cA$ is the infinitesimal gauge parameter. Obviously, 
$F^{}_{\mu\nu}$ and $R^{}_{\mu\nu}$ transform covariantly as well. Note that the 
matrix entries $\Xi^{a}{}^{}_{b}$ of the gauge parameter matrix $\Xi$ need not
be independent, see the Hermitian one--matrix model in Section~\ref{sec1mm} 
for a simple example.
In more complicated situations, it might not be possible to identify (or, for
other reasons, not desirable to work with) an independent set of gauge
generators. In that case one would have to work with a reducible gauge algebra,
and to introduce a new set of so--called stage--one gauge symmetries to handle 
the over--complete set of original gauge symmetries. In the BRST language this
leads to ghosts--for--ghosts. {}For a simple example of a stage--one reducible
gauge theory, see next Section~\ref{secconneslott}. Nevertheless, we shall for
the rest of this Section~\ref{sectargetspace} for simplicity assume that it is
possible to consistently pick an independent set of gauge parameters. It is
then possible to encode the gauge symmetry \e{gaugesym1} in a Fermionic
nilpotent BRST operator $\brst$ of the form 
\beq
\brst X^{}_{\mu}=(-1)^{\eps_{\mu}}[X^{}_{\mu}, C] = -[C,X^{}_{\mu}] , \qquad
\brst C  = -\Hf[ C, C] .  \label{brstsym1}
\eeq 
Here $C\!\in\!\cA$ is the target space ghost. The BRST operator $\brst$ is
by definition extended to polynomials in $X^{}_{\mu}$ and $C$ via a \nc Leibniz
rule,
\beq
\brst (f g) = (\brst f) g + (-1)^{\eps_{f}} f (\brst g) .
\eeq
In other words, the BRST operator ``$\brst$'' is a Fermionic vector field on a
\nc space. The square $\brst^2\!=\!\Hf[\brst,\brst]$ of the BRST operator is
again a vector field, which satisfies a \nc Leibniz rule
$\brst^{2}(f g)\!=\!(\brst^{2}f) g\!+\!f(\brst^{2}g)$, and is in fact identical
to zero,
\beq 
\brst^{2}=0. 
\eeq

\section{BV Odd Laplacian and Antibracket}
\label{secoddlapl}

\noi
The BRST formulation can be further encoded into the BV formalism
\cite{Batalin:1981jr,Batalin:1984jr}. If the gauge transformations form a
reducible or an open gauge algebra, this step will often be necessary. The
original BV recipe (which is formulated in terms of supercommutative field
variables $\phi^{\alpha}(x)$ in a path integral setting) can be directly
applied without modifications to \nc fields $\Phi^{\alpha}$ (where
$\Phi^{\alpha}$ is a collective notation for all fields
$\Phi^{\alpha}\!=\!\{X^{}_{\mu}, C,\ldots\}$) simply by treating the matrix
entries $(\Phi^{\alpha})^{a}{}^{}_{b}$ (which are supercommutative objects!) as
the fundamental variables. {}For instance, the odd Laplacian is
\beq
\Delta:=(-1)^{\eps((\Phi^{\alpha})^{a}{}^{}_{b})}
\papal{[(\Phi^{\alpha})^{a}{}^{}_{b}]}
\papal{[(\Phi^{*}_{\alpha})^{b}{}^{}_{a}]} ,
\eeq
where $\Phi^{*}_{\alpha}$ are the corresponding matrix--valued antifields. 
(We assume for simplicity that the matrices $\Phi^{\alpha}$ are world volume 
zero--forms.) The antibracket reads
\beq
(F,G):= F(\papar{[(\Phi^{\alpha})^{a}{}^{}_{b}]}
\papal{[(\Phi^{*}_{\alpha})^{b}{}^{}_{a}]}
-\papar{[(\Phi^{*}_{\alpha})^{a}{}^{}_{b}]}
\papal{[(\Phi^{\alpha})^{b}{}^{}_{a}]})G . 
\eeq
In particular, the antibrackets of fundamental variables read
\beq
\left((\Phi^{\alpha})^{a}{}^{}_{b}, (\Phi^{*}_{\beta})^{c}{}^{}_{d}\right)
=\delta^{\alpha}_{\beta}\delta^{a}_{d} \delta^{c}_{b} , \qquad
\left((\Phi^{\alpha})^{a}{}^{}_{b}, (\Phi^{\beta})^{c}{}^{}_{d}\right) = 0 , 
\qquad
\left((\Phi^{*}_{\alpha})^{a}{}^{}_{b}, (\Phi^{*}_{\beta})^{c}{}^{}_{d}\right)
 = 0 .
\eeq
Let us mention that the set ${\cal M}$ of points
$\Gamma^{A}\!\equiv\!(\Phi^{\alpha};\Phi^{*}_{\alpha})$ is called the
{\em antisymplectic phase space}. The antibracket $(\cdot,\cdot)$ is an 
antisymplectic structure on this phase space ${\cal M}$.

\noi
{\sc Remark}:
If one draws the index structure of a trace as a loop, then the antibracket
$(F,G)$ always joints two index loops $F\!=\!\tr f(\Phi,\Phi^{*})$ and
$G\!=\!\tr g(\Phi,\Phi^{*})$ into a single index loop. The action of the
antibracket $(\cdot,\cdot)$ on multiple loops can be determined via Leibniz
rule
\beq
(FG,H)=F(G,H)+(-1)^{\eps_{F}\eps_{G}+p^{}_{F}p^{}_{G}}G(F,H),
\eeq
so that in general
\beq
(\underbrace{\bigcirc\bigcirc\cdots\bigcirc}_{n~{\rm loops}},
\underbrace{\bigcirc\bigcirc\cdots\bigcirc}_{m~{\rm loops}})
=\sum \underbrace{\bigcirc\bigcirc\cdots\bigcirc\bigcirc\cdots\bigcirc
}_{n+m-1~{\rm loops}} .
\eeq
The odd Laplacian $\Delta$ adds an extra index loop $\Delta F$ when applied
to a single trace $F\!=\!\tr f(\Phi,\Phi^{*})$,
\beq
\Delta(\underbrace{\bigcirc}_{1~{\rm loop}}) 
= \sum\underbrace{\bigcirc \bigcirc}_{2~{\rm loops}} .
\eeq
The action of $\Delta$ on multiple loops can be determined from the formula
\beq
\Delta(FG) = (\Delta F)G +(-1)^{\eps_F}(F,G)+(-1)^{\eps_F} F(\Delta G),
\eeq
so that in general for $n\geq 2$,
\beq
\Delta(\underbrace{\bigcirc\bigcirc\cdots\bigcirc}_{n~{\rm loops}}) 
= \sum \underbrace{\bigcirc\bigcirc\cdots\bigcirc\bigcirc}_{n+1~{\rm loops}}
+\sum \underbrace{\bigcirc\cdots\bigcirc}_{n-1~{\rm loops}} .
\eeq
This picture superficially resembles the loop operator of Chas--Sullivan
in string topology \cite{chassullivan}, and the handle operator of Zwiebach
in closed string field theory \cite{Zwiebach:1992ie}, mostly because all the 
mentioned cases are governed by their underlying \BV algebras.

\section{BV Proper Action}
\label{secproper}

\noi
In the BV scheme \cite{Batalin:1981jr,Batalin:1984jr} one searches for a 
proper action $S$ to the classical master equation
\beq
 (S,S)=0 . \label{cme}
\eeq
In the above class of models, the minimal proper master action $S$ is given by
$S \!=\! \tr L$, where the Lagrangian density $L$ is
\beq
L = L^{}_{0} + (-1)^{\eps_{\mu}} X^{\mu *} \brst X^{}_{\mu} - C^{*} \brst C
\approx L^{}_{0} - (\brst X^{}_{\mu}) X^{\mu *} - (\brst C) C^{*} , 
\label{lagrdens1}
\eeq
and where $X^{\mu*}\!\in\!\cA$ and $C^{*}\!\in\!\cA$ are the corresponding
antifields, and ``$\approx$'' means equality modulo total commutator terms.
The antifields are generators of BRST symmetry. The classical BRST operator in
the BV formalism is $\brst\!=\!(S,\cdot)$. In general, there could be quantum
corrections to the classical master action $S$. However, quantum corrections
are not needed if $\Delta S\!=\!0$, which is true for the action \e{lagrdens1}.

\noi
{\sc Remark}:
Note that the BRST operator ``$\brst$'' acts on a whole matrix $\Phi^{\alpha}$ 
versus a matrix entry $(\Phi^{\alpha})^{a}{}^{}_{b}$ according to the rule
\beq
\brst[(\Phi^{\alpha})^{a}{}^{}_{b}]
= (-1)^{\eps_{a}}(\brst \Phi^{\alpha})^{a}{}^{}_{b} . \label{subtlefact}
\eeq

\noi
This sign factor \e{subtlefact} is due to a permutation of the row--index 
``$a$'' and BRST operator ``$\brst$''. (Recall that the matrix entries
$(\Phi^{\alpha})^{a}{}^{}_{b}$ of a supermatrix $\Phi^{\alpha}$ should strictly 
speaking be written as ${}^{a}(\Phi^{\alpha})^{}_{b}$.)
{}For a similar reason, if one identifies $\delta\leftrightarrow\mu\brst$ and 
$\Xi\leftrightarrow\mu C$ in \eqs{gaugesym1}{brstsym1}, where $\mu$ is a 
Fermionic parameter, then the matrix entries should be identified as
$\Xi^{a}{}^{}_{b}\leftrightarrow (-1)^{\eps_{a}}\mu C^{a}{}^{}_{b}$.

\section{BV Gauge--Fixing}
\label{secgf}

\noi 
The standard BV procedure to gauge--fix is to extend the Lagrangian density 
$L$ with a non--minimal sector $L \to L + \bar{C}^{*}\Pi$, where 
$\bar{C}\!\in\!\cA$ is an antighost and $\Pi\!\in\!\cA$ is a Lagrange 
multiplier, and $\bar{C}^{*},\Pi^{*}\in\cA$ are the corresponding antifields. 
In the end, all the antifields $\Phi^{*}_{\alpha}$ are replaced
\beq 
(\Phi^{*}_{\alpha})^{a}{}^{}_{b} \longrightarrow \frac{\partial \Psi}
{\partial[ (\Phi^{\alpha})^{b}{}^{}_{a}]} , \label{gaugefixantifields} 
\eeq 
where $\Psi\!=\!\Psi(\Phi)$ is a gauge fermion. It was proved in the original
work \cite{Batalin:1981jr,Batalin:1984jr} that the partition function
${\cal Z}$ is perturbatively well--defined and will locally not depend on the
gauge--fermion $\Psi$ as long as the quantum master equation holds, and the 
action and $\Psi$ satisfies certain rank conditions. Usually $\Psi$ is taken 
of the form
\beq 
\Psi = \tr\left(\bar{C}\chi\right) , \label{usualpsi}
\eeq
where $\chi\in\cA$ is the gauge--fixing condition. One possible gauge is a 
Lorenz type gauge 
\beq 
\chi = [n^{\mu}, X^{}_{\mu}] , \label{lorenzgauge} 
\eeq 
where $n^{\mu}\!\in\!\cA$ is a fixed vector. Gauge--fixing can be
considerably generalized, see \Ref{Batalin:2005my}.

\section{Star Product}
\label{secstarproduct}

\noi
Instead of matrices, it is also popular to formulate \nc field theories in
terms of fields $\phi^{\alpha}(x)$ (so-called symbols) and an associative star 
product ``$*$'', which is often taken to be of the Groenewold--Moyal type 
\beq
(f*g)(x)
= f(x)\exp\left[\papar{x^{}_{\mu}}m^{}_{\mu\nu}\papal{x^{}_{\nu}}\right] g(x) .
\label{moyal}
\eeq
The Groenewold--Moyal star product \e{moyal} corresponds to the case, where the
structure constants in \eq{xxfx1} yield a Heisenberg algebra,
\beq
f^{}_{\mu\nu}{}^{\lambda}=0 ,  \qquad 
\theta^{}_{\mu\nu} = m^{}_{\mu\nu} - (-1)^{\eps_{\mu}\eps_{\nu}}m^{}_{\nu\mu} .
\eeq
\BV formalism also works in this setting \cite{Barnich:2001mc,Barnich:2002tz}
(since the symbols are supercommutative!), and considerations of {\em local} 
BRST cohomology \cite{Barnich:1994db} have been extended to \nc field theories
\cite{Barnich:2003wq}, at least when using the pragmatic definition of
locality. The pragmatic definition of a local functional
\beq 
{}F = \int\! dx\ f(x)  \label{fint}
\eeq 
is an integral over a function 
\beq 
f(x)=f(\phi(x), \partial\phi(x), \ldots, \partial^{N}\phi(x),x) \label{ffct}
\eeq 
that depends locally on the fields $\phi^{\alpha}(x)$ in the point $x$ and 
its derivatives to some finite order $N$. The corresponding definition of a 
local functional $F$ in a matrix--setting is, roughly speaking, a single--trace
\beq 
{}F = \tr f(\Phi) , 
\eeq 
where $f\!=\!f(\Phi)$ is a polynomial in the $\Phi^{\alpha}$'s. It could be
interesting to investigate local BRST cohomology from this 
matrix--point--of--view.

\section{Hermitian One--Matrix Model}
\label{sec1mm}

\noi
Consider a Hermitian one--matrix model
$L^{}_{0}(H)=\sum_{n}a^{}_{n} H^{n}/n!$, where $X\!=\!H$ is a Bosonic Hermitian 
endomorphism in a $(N^{}_{0}|N^{}_{1})$ super vector space $V$ of dimension
$N\!=\!N^{}_{0}\!+\!N^{}_{1}$, and where the $a^{}_{n}$'s are Bosonic numbers. 
The original action $S^{}_{0}=\tr L^{}_{0}$ is 
invariant under gauge transformations $H\to H^{g}\!=\!g^{-1}Hg$, where
$g=e^{\Xi}\!\in\!U(N^{}_{0}|N^{}_{1})$. The model has $N^{2}$ gauge parameters
$\Xi^{a}{}^{}_{b}$ corresponding to the number of matrix entries in $H$. 
However, the Bosonic eigenvalues $\lambda^{}_{1}$, $\lambda^{}_{2}$, $\ldots$, 
$\lambda^{}_{N}$, of $X$ are $N$ gauge--invariant quantities, which cannot be 
changed by gauge transformations of adjoint type. Hence there are actually
only $N(N\!-\!1)$ independent gauge parameters. Thus the gauge algebra is
reducible.

\noi 
{}For a diagonal matrix $H$, the $N$ redundant gauge parameters may be 
identified with the diagonal matrix entries $\Xi^{1}{}^{}_{1}$,
$\Xi^{2}{}^{}_{2}$, $\ldots$, $\Xi^{N}{}^{}_{N}$, at the infinitesimal level. 
It is possible to truncate the reducible gauge algebra to a stage--zero
irreducible gauge algebra as follows. Since all Hermitian matrices $H$ are
diagonalizable, it is always possible to pick a diagonal gauge. We implement
the diagonal gauge via a Lorenz type gauge condition
\beq 
\chi = [n,H] , \qquad  n=\diag(\nu^{}_{1},\ldots,\nu^{}_{N}) , \qquad
\chi^{a}{}^{}_{b}=(\nu^{}_{a}\!-\!\nu^{}_{b})H^{a}{}^{}_{b} , 
\label{diaggauge}
\eeq 
where $n$ is a fixed diagonal matrix with different eigenvalues $\nu^{}_{1}$,
$\nu^{}_{2}$, $\ldots$, $\nu^{}_{N}$. Since there are only $N(N\!-\!1)$
independent gauge symmetries, the ghost $C$ and antighost $\bar{C}$ have only
off-diagonal entries. There are also only be $N(N\!-\!1)$ off-diagonal
$\chi^{a}{}^{}_{b}$ gauge conditions \e{diaggauge} to implement,
$1\leq a \neq b\leq N$, so the Lagrange multiplier $\Pi$ contains only
off-diagonal entries as well.

\noi
The antifields $\Phi^{*} \!=\! \partial \Psi / \partial \Phi$ with 
$\Psi\!=\!\tr\left(\bar{C}\chi\right)$ become
\beq
H^{*}=[\bar{C},n], \qquad C^{*}= 0 , \qquad
\bar{C}^{*}= \chi , \qquad \Pi^{*}=0 .
\eeq

\noi
The gauge--fixed action \e{lagrdens1} reads

\begin{center}
\begin{tabular}{cccccccc}
$\left. S \right|^{}_{\Phi^{*} = \frac{\partial \Psi}{ \partial \Phi}}$
&$\sim$&$S^{}_{0}$&$+$&$\tr\left(\bar{C}[n,[H,C]]\right)$&$+$&
$\tr\left([n,H]\rule[-1.5ex]{0ex}{4.5ex} \right.$&
$\left.\rule[-1.5ex]{0ex}{4.5ex} \Pi\right)$ \\
&$\sim$&$S^{}_{0}$&$+$&
$\sum_{a\neq b}(-1)^{\eps_{b}} \left[
\bar{C}^{b}{}^{}_{a}(\nu^{}_{a}\!-\!\nu^{}_{b})(H^{a}{}^{}_{a}\!-\!H^{b}{}^{}_{b})
C^{a}{}^{}_{b} \rule[-1.5ex]{0ex}{4.5ex} \right.$ &
$+$&$(\nu^{}_{b}\!-\!\nu^{}_{a})H^{b}{}^{}_{a}$&
$\left.\rule[-1.5ex]{0ex}{4.5ex} \Pi^{a}{}^{}_{b}\right]$. \\
{\rm Gauge}-&&{\rm Orig.}&&{\rm Faddeev}-&&{\rm Gauge}&{\rm Lagr.} \\
{\rm fixed}&&{\rm ac-}&&{\rm Popov}&&{\rm condi-}&{\rm mult.} \\
{\rm action}&&{\rm tion}&&{\rm matrix}&&{\rm tion}&
\end{tabular}
\end{center}

\noi
The partition function ${\cal Z}$ becomes
\bea
{\cal Z} &=& \int \! [dH][dC][d\bar{C}][d\Pi] \ 
e^{\ih S(\Phi,\Phi^{*}=\frac{\partial \Psi}{\partial\Phi} )} \cr
&\sim& \int \! d\lambda^{}_{1} \cdots d\lambda^{}_{N} \
e^{\ih \tr L^{}_{0}\left(\diag(\lambda^{}_{1},\ldots,\lambda^{}_{N})\right)}
\Delta^{2}(\lambda^{}_{a})
\label{zresult}
\eea
up to a numerical factor, where the super-Vandermonde determinant is 
\beq
\Delta(\lambda^{}_{a})=
\prod_{1\leq a<b\leq N}
(\lambda^{}_{b}\!-\!\lambda^{}_{a})^{\left[(-1)^{\eps_{a}+\eps_{b}}\right]} .
\label{supervandermonde}
\eeq
The result \e{zresult} is manifestly
independent of the gauge--fixing parameters $\nu^{}_{1}$, $\nu^{}_{2}$,
$\ldots$, $\nu^{}_{N}$, as it should be. The integrand consists of a classical
Boltzmann factor times a square $\Delta^{2}(\lambda^{}_{a})$ of a Vandermonde
superdeterminant, whose $N(N\!-\!1)$ factors reflect the $N(N\!-\!1)$ 
independent gauge symmetries.

\noi
The above removal of the $N$ diagonal gauge parameters directions
$\Xi^{a}{}^{}_{a}\!=\!0$, $a\!\in\!\{1,2,\ldots, N\}$, can also be seen as a way
to get rid of $N$ zero--modes in the Faddeev--Popov determinant (if one assumes
that all the eigenvalues $\lambda^{}_{1}$, $\lambda^{}_{2}$, 
$\ldots$, $\lambda^{}_{N}$, are different).

\section{Connes--Lott Model for a $2$--Point Space}
\label{secconneslott}

\noi
The algebra $\cA\!=\!{\rm End}(V)$ of the Connes--Lott model
\cite{Connes:1990qp} consists of endomorphisms in a $(1|1)$ super vector space
$V$, \ie the vector space $V$ has one Bosonic and one Fermionic direction. One
may think of the endomorphisms as $2\!\times\!2$ matrices. We will for
simplicity only consider matrices that are either diagonal or off--diagonal and
that carry definite Grassmann--parity. Note that diagonal and off--diagonal
matrices (with matrix entries of the same Grassmann--parity) carry opposite
Grassmann--parity.

\noi 
The Connes--Lott model for a $2$--point space has only one algebra generator
$x^{}_{1}$ and one covariant coordinate $X^{}_{1}\!=\!x^{}_{1}\!+\!A^{}_{1}$. 
They are off--diagonal Fermionic matrices 
\beq 
x^{}_{1}=\twobytwo{0}{1}{1}{0}, \qquad X^{}_{1}=\twobytwo{0}{H}{\bar{H}}{0} , 
\eeq 
where $H$ is a complex--valued Bosonic Higgs field, and $\bar{H}$ is the
complex conjugated field. The single world volume coordinate $x^{}_{1}$ is a
\nc coordinate, 
\beq
[x^{}_{1},x^{}_{1}]=\theta^{}_{11} {\bf 1} , \qquad \theta^{}_{11} = 2 .
\eeq
The original action $S^{}_{0}=\tr L^{}_{0}$ is given as
\beq
L^{}_{0}\sim F^{}_{11}F^{11}\Gamma
=\frac{1}{4} (F^{}_{11})^{2}\Gamma 
= (|H|^{2}\!-\!1)^{2}\Gamma, 
\eeq
where
\beq
 F^{}_{11}=[x^{}_{1},A^{}_{1}]+[A^{}_{1},x^{}_{1}]+[A^{}_{1},A^{}_{1}]
=[X^{}_{1},X^{}_{1}]-[x^{}_{1},x^{}_{1}] = 2(|H|^{2}\!-\!1) {\bf 1} ,
\eeq
and where $\Gamma$ is a chirality operator,
\beq
\Gamma:=\twobytwo{1}{0}{0}{-1} .
\eeq
The chirality operator $\Gamma$ breaks down a $U(1|1)$ supergroup (which 
naturally acts on the $(1|1)$ vector space $V$) to a diagonal $U(1)\times U(1)$ 
subgroup. In detail, the gauge group element $g \in U(1)\times U(1)$ is of
the form
\beq 
g=e^{i\Xi}=\twobytwo{e^{i\xi}}{0}{0}{e^{i\xi'}} ,
\eeq 
with gauge parameter 
\beq 
\Xi=\twobytwo{\xi}{0}{0}{\xi'} . 
\eeq 
The transformed covariant coordinate $X_{1}^{g}$ is 
\beq 
X_{1}^{g}=g^{-1}X^{}_{1}g = \twobytwo{0}{H^{g}}{\bar{H}^{g}}{0} , \qquad 
H^{g}=H e^{-i\xi^{}_{-}}  , \qquad  \xi^{}_{\pm}:=\xi\!\pm\!\xi' .
\eeq 
The eigenvalues $\pm |H|$ of the matrix $X^{}_{1}$ (and hence the modulus 
$|H|$) 
are preserved under gauge transformations, because they are just similarity 
transformations. The infinitesimal gauge transformation reads 
\beq 
\delta X^{}_{1} = i [X^{}_{1}, \Xi] , \qquad 
\delta(\Re(H)) = \xi^{}_{-} \Im(H) ,  \qquad 
\delta(\Im(H)) = -\xi^{}_{-} \Re(H) . 
\eeq
Clearly, the two $U(1)$ gauge factors are linearly dependent, \ie they 
constitute a reducible gauge algebra. The gauge--for--gauge symmetry
$\tilde{\delta}$ is of the form
\beq
\tilde{\delta}\Xi =\twobytwo{\tilde{\delta}\xi}{0}{0}{\tilde{\delta}\xi'}
= \twobytwo{1}{0}{0}{1} \zeta , \qquad 
\tilde{\delta}\xi^{}_{+}=2\zeta , \qquad 
\tilde{\delta}\xi^{}_{-}=0 ,
\eeq
where $\zeta$ is a gauge--for--gauge parameter. Although it is immediately
clear that we can go to an irreducible basis by fixing $\xi^{}_{+}\!=\!0$, 
let us here for illustrative purposes show how to treat the Connes--Lott 
$2$-point model as a stage--one reducible gauge system \cite{Batalin:1984jr}.
\footnote{We should mention \Ref{Huffel:2001ea} that also applies the BV recipe
to the Connes--Lott $2$-point model. (See \Ref{Huffel:2002au} for a review of 
\Ref{Huffel:2001ea}.) The method of \Ref{Huffel:2001ea} (implicitly) requires 
that all higher--stage fields should be $2\!\times\!2$ matrix--valued, and as a
consequence, ends up with infinitely many reducibility stages by alternatingly
overshooting and undershooting the single gauge--symmetry similar to the
alternating series $1\!-\!2\!+\!2\!-\!2\!+\!2\ldots$. Such infinite tower of
fields is ill--defined and plagued with anomalies, \ie the resulting partition
function ${\cal Z}$ will depend on the choice of the gauge--fixing condition.
It would be out of scope to show this in detail here, but the quickest argument
is probably to notice that the final formula for the gauge--fixed action (after
the infinitely many higher--stage fields have been heuristically integrated 
out; see formula (4.8) in \Ref{Huffel:2001ea}, or equivalently, formula (14) in
\Ref{Huffel:2002au}) contains {\em two} Faddeev--Popov ghost--antighost pairs 
but only {\em one} independent gauge--condition. Recall that in the usual
Faddeev--Popov approach, the number of gauge--conditions must precisely match
the number of ghost--antighost pairs. Arguments along these lines show that the
method of \Ref{Huffel:2001ea} will depend on the gauge--fixing choice, and the 
method therefore produces a useless result. We shall here avoid the same 
ill--fate by allowing for $1\!\times\!1$ matrix--valued stage--one fields.} 

\noi
The Fermionic reducible ghost is
\beq
C=\twobytwo{c}{0}{0}{c'} .
\eeq
The BRST transformations are
\bea
\twobytwo{0}{\brst H }{- \brst \bar{H}}{0} 
\equi{\e{subtlefact}} \brst X^{}_{1} &=& - i [X^{}_{1}, C]
=-i\twobytwo{0}{ H c^{}_{+} }{ \bar{H}c^{}_{+}}{0} , \\
\brst(\Re(H)) =  c^{}_{+} \Im(H) , &&  
\brst(\Im(H)) = - c^{}_{+} \Re(H) ,\qquad
c^{}_{\pm}:=c\!\pm\!c' ,  \\
\twobytwo{\brst c}{0}{0}{-\brst c'} \equi{\e{subtlefact}}
\brst C &=& \twobytwo{1}{0}{0}{1} \eta ,
\qquad \brst c^{}_{+} = 0 , \qquad \brst c^{}_{-} = 2\eta ,
\eea
where $\eta$ is a Bosonic ghost--for--ghost. Nilpotency imposes
$\brst\eta\!=\!0$.

\noi
{\sc Remark}: If one identifies $\delta\leftrightarrow\mu\brst$, 
$\tilde{\delta}\leftrightarrow\tilde{\mu}\brst$, and $\Xi\leftrightarrow\mu C$,
where $\mu$ and $\tilde{\mu}$ are Fermionic parameters, then one should
identify $\xi^{}_{\mp} \leftrightarrow \mu c^{}_{\pm}$ and
$\zeta\leftrightarrow \mu\tilde{\mu}\eta$.

\noi
In the non--minimal sector, the antighost $\bar{C}$ and the Lagrange multiplier
$\Pi$ are 
\beq
\bar{C}=\twobytwo{\bar{c}}{0}{0}{\bar{c}'},  \qquad
\Pi=\twobytwo{\pi}{0}{0}{\pi'}. 
\eeq 
One also has to introduce an antighost--for--ghost $\bar{\eta}$ and a
Lagrange--multiplier--for--ghost $\bar{\pi}$. Moreover, there are an extra 
ghost $\tilde{\eta}$ and an extra Lagrange multiplier $\tilde{\pi}$. And
finally, all the fields have corresponding antifields.

\noi
A proper stage--one reducible master action $S$ is
\beq
S=S^{}_{0}+ \tr\left(-X^{1*} \brst X^{}_{1} -  C^{*} \brst C 
+ \bar{C}^{*}\Pi\right)
+ \bar{\eta}^{*} \bar{\pi} + \tilde{\eta}^{*}\tilde{\pi} .
\eeq
A suitable gauge--fermion $\Psi$ can be chosen on the form
\beq
\Psi = \tr\left(\bar{C}\chi\right) + \bar{\eta} \tr\left(\Gamma C\right)
+ \tr\left(\bar{C}\Gamma\right) \tilde{\eta} . \label{clpsi1}
\eeq
The fixed one-dimensional Fermionic vector $n^{1}$ from \eq{lorenzgauge} can
be chosen as
\beq
n^{1}=\twobytwo{0}{e^{i\theta}}{e^{-i\theta}}{0},
\eeq
where $\theta$ is an angle. The Lorenz type gauge condition $\chi$ reads
\beq
\chi = [n^{1},X^{}_{1}]+\alpha\Gamma\Pi 
= 2\Re(H e^{-i\theta}){\bf 1}+\alpha\Gamma\Pi ,
\eeq
where $\alpha$ is a gauge--fixing parameter. Singular (\ie 
delta--function--type) gauge--fixing corresponds to $\alpha\!=\!0$, while 
Gaussian--type gauge--fixing corresponds to $\alpha\!\neq\!0$.
Hence the gauge--fermion $\Psi$ from \eq{clpsi1} takes the form
\beq
\Psi = \bar{c}^{}_{+}\left[ 2 \Re(H e^{-i\theta})
+\frac{\alpha}{2}\pi^{}_{-}\right] + \bar{\eta}c^{}_{-}
+ \bar{c}^{}_{-}\left[ \tilde{\eta}+\frac{\alpha}{2}\pi^{}_{+}\right] , 
\eeq
where $\bar{c}^{}_{\pm}:=\bar{c}\!\pm\!\bar{c}'$ and 
$\pi^{}_{\pm}:=\pi\!\pm\!\pi'$. 

\noi
The antifields $\Phi^{*} \!=\! \partial \Psi / \partial \Phi$ become
\beq
X^{1*}=[\bar{C},n^{1}], \qquad C^{*}= \bar{\eta} \Gamma , \qquad
\bar{C}^{*}= \chi + \Gamma \tilde{\eta} , \qquad
\Pi^{*}=\alpha\bar{C}\Gamma , \qquad
\bar{\eta}^{*}=c^{}_{-} , \qquad \tilde{\eta}^{*}=\bar{c}^{}_{-} ,
\eeq
and all the remaining antifields $\eta^{*}$,  $\bar{\pi}^{*}$, and
$\tilde{\pi}^{*}$ are zero.

\noi
The gauge--fixed stage--one reducible action reads
\beq
\left. S \right|_{\Phi^{*} = \partial \Psi / \partial \Phi}
=S^{}_{0} +\bar{c}^{}_{+} 2\Im(He^{-i\theta}) c^{}_{+}  - 2\bar{\eta} \eta +
\left[2\Re(H e^{-i\theta})+\frac{\alpha}{2}\pi^{}_{-}\right]\pi^{}_{-} 
+\left[ \tilde{\eta}+\frac{\alpha}{2}\pi^{}_{+}\right] \pi^{}_{+} 
+ c^{}_{-} \bar{\pi} + \bar{c}^{}_{-} \tilde{\pi} .
\eeq
If one integrates over $\bar{\eta}$, 
$\eta$, $\tilde{\eta}$, $\pi^{}_{+}$, $c^{}_{-}$, $\bar{\pi}$, 
$\bar{c}^{}_{-}$, and $\tilde{\pi}$ in the path integral, one arrives at the 
standard gauge--fixed stage--zero irreducible action

\begin{center}
\begin{tabular}{cccccccc}
$\left. S \right|_{\Phi^{*} = \partial \Psi / \partial \Phi}$
&$\sim$&$S^{}_{0} $&$+$&$\bar{c}^{}_{+} 2\Im(He^{-i\theta}) c^{}_{+}$&$+$&
$\left[2 \Re(H e^{-i\theta})+\frac{\alpha}{2}\pi^{}_{-}\right]$&$\pi^{}_{-}$, \\
{\rm Gauge}-{\rm fixed}&&{\rm Original}&&
{\rm Faddeev}-{\rm Popov}&&{\rm Gauge}&{\rm Lagr.} \\
{\rm action}&&{\rm action}&&$1\!\times\!1$ \ {\rm matrix}&&{\rm cond.}&
{\rm mult.}
\end{tabular}
\end{center}

\noi
with the remaining field content $H$, $c^{}_{+}$, $\bar{c}^{}_{+}$, and
$\pi^{}_{-}$. The Lagrange multiplier $\pi^{}_{-}$ gauge--fixes in the singular
limit $\alpha\!=\!0$ the Higgs field $H$ to two opposite values
$H\!=\!\pm|H|e^{i\theta}$. Here we encounter a technical (as opposed to a 
fundamental) Gribov ambiguity, since our simple type of gauge condition $\chi$
picks a line through the origin, which always will intersect the gauge orbit
(=circle) in precisely two opposite points. (Clearly, at the fundamental level,
one should just find a gauge condition that picks a half--line instead,
although we shall not implement this in practice here, since it is anyway not
needed.)

\section{Conclusions}
\label{secconcl}

\begin{itemize}
\item
We have, first of all, seen that the \BV formalism \cite{Batalin:1981jr} is a 
useful tool to gauge--fix matrix models, or \nc field theories, since such 
theories may exhibit reducible gauge symmetries. 

\item
We have for the first time shown how to successfully treat the 
Connes--Lott model \cite{Connes:1990qp} within the reducible \BV framework 
\cite{Batalin:1984jr}, \cf Section~\ref{secconneslott}. 

\item
When considering matrix models one inevitable faces Itzykson--Zuber--like 
integrals. We have for the first time {\em explicitly} demonstrated the 
localization mechanism for the $U(N_{0}|N_{1})$ 
Harish--Chandra--Itzykson--Zuber (HCIZ) integral \cite{hc1957,Itzykson:1979fi},
\cf Appendix~\ref{seciz}. By the word {\em explicitly}, we mean, in particular,
that we do {\em not} rely on the Duistermaat--Heckman Localization Theorem
\cite{Duistermaat:1982vw}. 
\end{itemize}

\vspace{0.8cm}

\noi {\sc Acknowledgement:}~K.B.\ would like to thank Igor Batalin for
discussions, and both the University of Vienna and the Erwin Schr\"odinger
Institute for warm hospitality. The work of K.B.\ is supported by the Ministry
of Education of the Czech Republic under the project MSM 0021622409.

\appendix

\section{HCIZ Integrals and Localization}
\label{seciz}

\noi
Let $\cA\!=\!{\rm End}(V)$ be the algebra of endomorphisms in a 
$(N^{}_{0}|N^{}_{1})$ super vector space $V$ of dimension 
$N\!=\!N^{}_{0}\!+\!N^{}_{1}$. Consider the 
Harish--Chandra--Itzykson--Zuber (HCIZ) integral \cite{hc1957,Itzykson:1979fi}
\beq
{\rm HCIZ}(A,B) = \int^{}_{U\in U(V)} \!\!\!\!\!\!\!\!
\rho(U) dU e^{\ih S^{}_{0}} , 
\qquad S^{}_{0}=\tr L^{}_{0} , \qquad L^{}_{0} = AUBU^{\dag} , \label{izint0}
\eeq
where the integration variable 
$U\!\in\!U(N_{0}|N_{1})\!\equiv\!U(V)\!\subseteq\!\cA$ is a unitary
endomorphism, $\eps(U)\!=\!0$, and where $A,B\!\in\!\cA$ are two fixed Bosonic 
Hermitian matrices, $\eps(A)\!=\!0\!=\!\eps(B)$. This integral is, \eg of 
great importance in solving two--matrix--models. Let us choose a basis for $V$.
The Haar measure is
\beq
\int^{}_{U\in U(V)}\!\!\!\!\!\!\!\! \rho(U)dU\ldots 
\sim \int^{}_{U\in{\rm End}(V)} \!\!\!\!\!\!\!\! [dU][dU^{\dag}]
\delta(U^{\dag}U\!-\!{\bf 1})\ldots
\sim \int^{}_{U\in{\rm End}(V)} \!\!\!\!\!\!\!\! [dU][dU^{\dag}][d\Pi] 
e^{\ih \tr\left((U^{\dag}U-{\bf 1})\Pi\right)}\ldots ,
\eeq
where $\Pi\!\in\!\cA$ is an Bosonic Hermitian matrix that plays the r\^ole of 
Lagrange multiplier for the unitarity constraint $U^{\dag}U\!=\!{\bf 1}$. The 
Haar measure is invariant under the left--right action of $U(V)$, 
\beq
U \longrightarrow U^{}_{L} U U^{}_{R}, \qquad
\Pi \longrightarrow U^{\dag}_{R} \Pi U^{}_{R}, \qquad 
U^{}_{L} , U^{}_{R} \in U(V), \qquad U \in{\rm End}(V) . 
\eeq
Hence we can (and will) assume without loss of generality that the fixed
matrices $A$ and $B$ are both diagonal matrices
\beq
A=\diag(\lambda^{}_{1},\lambda^{}_{2},\ldots, \lambda^{}_{N})  , \qquad 
B=\diag(\mu^{}_{1},\mu^{}_{2},\ldots, \mu^{}_{N}) , 
\eeq
with Grassmann--parity $\eps(\lambda^{}_{a})\!=\!0\!=\!\eps(\mu^{}_{a})$, 
$a\in\{1,2, \ldots, N\}$. In particular, $[A,B]\!=\!0$. We shall furthermore 
assume that the eigenvalues $\lambda^{}_{1}$, $\lambda^{}_{2}$, $\ldots$,
$\lambda^{}_{N}$ of the matrix $A$ are different, and similarly, that the
eigenvalues $\mu^{}_{1}$, $\mu^{}_{2}$, $\ldots$, $\mu^{}_{N}$ of the matrix $B$
are different. We want to prove a superversion of the 
Harish--Chandra--Itzykson--Zuber formula \cite{Alfaro:1994ca,Guhr:1996mx}
\beq
{\rm HCIZ}(A,B) = {\rm constant} \times \frac{
\det\left(e^{\ih\lambda^{}_{a}\mu^{}_{b}}
\right)^{}_{1\leq a,b\leq N^{}_{0}}}{\Delta(\lambda^{}_{a})}\frac{ 
\det\left(e^{-\ih\lambda^{}_{a}\mu^{}_{b}}
\right)^{}_{N^{}_{0}+1\leq a,b\leq N}}{\Delta(\mu^{}_{b})} 
\label{finalhcizformula}
\eeq
up to an overall numerical factor, which we ignore, since it is often
irrelevant in physics applications. The formula \e{finalhcizformula} coincides
with the one-loop approximation of the asymptotic steepest decent expansion 
for $\hbar\!\to\!0$, \cf \Ref{Stone:1988fu} and \Ref{Szabo:2000qq}. Our goal 
with this Appendix~\ref{seciz} is to provide a fully explicit localization 
argument that the one-loop approximation is the exact result. (In particular, 
we shall {\em not} rely on the Duistermaat--Heckman Localization Theorem 
\cite{Duistermaat:1982vw}, although our method is in principle equivalent.
Beware that many articles, that claim to use Duistermaat--Heckman Theorem to
prove localization, do actually not show that the assumptions in the 
Duistermaat--Heckman Theorem are fulfilled, and hence give incomplete
localization arguments.) The original derivations in \Ref{Alfaro:1994ca} and
\Ref{Guhr:1996mx} of formula \e{finalhcizformula} use superversions of the 
heat equation method and the Gelfand--Tzetlin coordinate approach, 
respectively.

\subsection{Instantons/Classical Solutions}

\noi
An infinitesimal variation $\delta S^{}_{0}$ of the original action 
$S^{}_{0}\!=\!\tr\left(AUBU^{-1}\right)$ reads
\beq
\delta S^{}_{0} = \tr\left( \EOM U^{-1} \delta U \right) , 
\eeq
with classical equations of motion
\beq 
\EOM := \left[B, H \right] , \qquad H := U^{-1} A U  .
\eeq
The classical equations of motion $\EOM\!\approx\!0$ implies that $H$ is 
diagonal, \ie there exists a permutation $\sigma\!\in\!S^{}_{N}$ such that
\beq
H \approx \diag\left(\lambda^{}_{\sigma(1)},\lambda^{}_{\sigma(2)},
\ldots, \lambda^{}_{\sigma(N)}\right)  \qquad \Leftrightarrow \qquad
\left(\lambda^{}_{a}\!-\!\lambda^{}_{\sigma(b)}\right) U^{a}{}^{}_{b}
\approx 0 , \label{hclassol}
\eeq
where ``$\approx$'' means equality modulo classical equations of motion. Thus
the matrix $U$ can at most have one non-zero entry  $U^{a}{}^{}_{\sigma(a)}$ in
each row ``$a$'' (and similarly at most one non-zero entry in each column). 
On the other hand, to ensure that the matrix $U$ is invertible, all the entries
of the form $U^{a}{}^{}_{\sigma(a)}$ must be non-zero and Bosonic. This is
precisely possible if the permutation $\sigma\!\in\!S^{}_{N}$ does not mix
Bosonic and Fermionic directions in $V$, \ie 
$\sigma\!\in\!S^{}_{N^{}_{0}}\!\times\!S^{}_{N^{}_{1}}$. Let 
\beq
S^{}_{N^{}_{0}} \times S^{}_{N^{}_{1}} \ni \sigma
\mapsto P^{}_{\sigma} \in U(N_{0}|N_{1})\equiv U(V) \subseteq \cA 
\eeq
denote the canonical embedding
$S^{}_{N^{}_{0}} \times S^{}_{N^{}_{1}} \to U(N_{0}|N_{1})$.
The full classical solution for $U$ is a permutation matrix $P^{}_{\sigma}$ 
times an element $e^{i\Xi}$ of the Cartan torus, 
\beq
U \approx P^{}_{\sigma} e^{i\Xi} , \qquad 
\sigma\in S^{}_{N^{}_{0}}\!\times\!S^{}_{N^{}_{1}} , \qquad 
\Xi=\diag(\xi^{}_{1},\xi^{}_{2},\ldots, \xi^{}_{N}) . \label{uclassol} 
\eeq
The stationary surface of classical $U$-solutions is a disjoint union of 
instanton sectors, which are labelled by the permutations 
$\sigma\!\in\!S^{}_{N^{}_{0}}\!\times\!S^{}_{N^{}_{1}}$. 

\noi
The original action $S^{}_{0}$ has a $U(1)^{N}\!\times\!U(1)^{N}$ gauge symmetry
corresponding to the left and the right Cartan torus, 
\beq
U \to e^{i\Xi^{\prime}} U e^{i\Xi} , \qquad 
\Xi=\diag(\xi^{}_{1},\xi^{}_{2},\ldots, \xi^{}_{N}) , \qquad 
\Xi^{\prime}
=\diag(\xi^{\prime}_{1},\xi^{\prime}_{2},\ldots, \xi^{\prime}_{N}) .
\label{izgaugesym}
\eeq
Since the gauge group $U(1)^{N}\!\times\!U(1)^{N}$ is compact, gauge--fixing is
actually not necessary, and we shall ignore it.

\subsection{A Fermionic Symmetry ``$\brst$''}

\noi
In anticipation of at least one Vandermonde determinant in the final formula 
\e{finalhcizformula}, let us consider the partition function
\beq
{\cal Z}^{}_{1} = {\rm HCIZ}(A,B) \Delta(\mu^{}_{b})
=\int^{}_{U\in{\rm End}(V)} \!\!\!\!\!\!\!\!
[dU][dU^{\dag}][d\Pi][dC] e^{\ih S^{}_{1}} , \label{izint1} 
\eeq
where
\beq
S^{}_{1} = \tr L^{}_{1} , \qquad 
L^{}_{1} =  HB + \Hf C[B,C] + (U^{\dag}U\!-\!{\bf 1})\Pi , \qquad   
H := U^{\dag}AU . \label{ess1}
\eeq
The first, second, and third term in $L^{}_{1}$ implements the original HCIZ 
action $S^{}_{0}$, the Vandermonde determinant $\Delta(\mu^{}_{b})$, and the 
unitarity constraint $U^{\dag}U\!=\!{\bf 1}$, respectively. The $C\!\in\!\cA$
is an (anti)Hermitian and off--diagonal Fermionic matrix. In particular, its 
diagonal entries $C^{a}{}^{}_{a}\!=\!0$ are zeroes, $1\!\leq\!a\!\leq\!N$. 
(One should mention that the Gaussian Bosonic $C^{a}{}^{}_{b}$-integrations in 
\eq{izint1} are defined via analytic continuation, \ie one should integrate 
along a straight line through the origin in the complex $C^{a}{}^{}_{b}$-plane,
in such a way that the integrand becomes exponentially damped.)

\noi
To show that the integral ${\cal Z}^{}_{1}$ localizes on the classical
solutions \e{hclassol}, one uses a divergence--free Grassmann--odd left vector 
field ``$\brst$'', 
\bea
(-1)^{\eps_{a}}\brst (U^{a}{}^{}_{c}) \equiv (\brst U)^{a}{}^{}_{c} 
= U^{a}{}^{}_{b} C^{b}{}^{}_{c} , && 
(-1)^{\eps_{a}}\brst (C^{a}{}^{}_{b}) \equiv (\brst C)^{a}{}^{}_{b} 
=H^{\prime a}{}^{}_{b}  := \left\{\begin{array}{rcl} H^{a}{}^{}_{b} 
&\for&a\!\neq\!b , \cr 0 &\for&a\!=\!b , \end{array}\right.
\label{delta1a} \cr
(-1)^{\eps_{a}}\brst (U^{\dag}{}^{a}{}^{}_{c}) 
\equiv (\brst U^{\dag})^{a}{}^{}_{c} 
= - C^{a}{}^{}_{b} U^{\dag}{}^{b}{}^{}_{c} , &&  
(-1)^{\eps_{a}}\brst (\Pi^{a}{}^{}_{b}) \equiv (\brst\Pi)^{a}{}^{}_{b} 
=[\Pi,C]^{a}{}^{}_{b} . 
\label{delta1b}
\eea
{}For a review of localization techniques, see, \eg \Ref{Szabo:1996md}.
The left vector field ``$\brst$'' is by definition a linear derivation 
$\brst (f g) \!=\! (\brst f) g \!+\! (-1)^{\eps_{f}} f (\brst g)$.
The definition \e{delta1b} implies the following compact formulas
\beq
\brst U = U C , \qquad \brst U^{\dag} = -C U^{\dag}  , \qquad 
 \brst C = H^{\prime} , \qquad  \brst H = [H, C] , \qquad  
\brst \Pi = [\Pi, C] , \label{delta2}
\eeq 
where $H^{\prime}$ denotes the $H$-matrix with zeroes in the diagonal. Now
it turns out that the $S^{}_{1}$ action \e{ess1} is invariant under the 
Grassmann--odd $\brst$ vector field
\beq
\brst S^{}_{1} = \tr\left(B\brst H -  C[B,\brst C] + U\Pi\brst U^{\dag} 
+ \Pi U^{\dag}\brst U +(U^{\dag}U\!-\!{\bf 1})\brst\Pi \right) 
= \tr\left(B[H,C] - C[B,H^{\prime}] \right) = 0 .
\eeq

\subsection{Cohomology of $\brst$}

\noi
The divergence ${\rm div}(\brst)$ of the Fermionic vector field $\brst$ 
vanishes
\beq
{\rm div}(\brst) 
= \papal{U^{a}{}^{}_{b}}\brst( U^{a}{}^{}_{b}) 
+ \papal{U^{\dag}{}^{a}{}^{}_{b}}\brst( U^{\dag}{}^{a}{}^{}_{b}) 
+\papal{C^{a}{}^{}_{b}}\brst( C^{a}{}^{}_{b})
+\papal{\Pi^{a}{}^{}_{b}}\brst(\Pi^{a}{}^{}_{b}) = 0 , 
\label{divergencefree}
\eeq
\cf definition \e{delta1b}. (The underlying reason for zero divergence is the 
right invariance of the Haar measure $\rho(U)dU$.) Integration by part shows
that an integral $\int^{}_{U\in U(V)} \! \rho(U)dU[dC] \brst f\!=\!0$ over an 
$\brst$-exact quantity $\brst f$ is zero, if there are no boundary 
contributions. Here $f\!=\!f(U,C)$ is a function. (There are never boundary 
contributions from Fermionic integrations, nor from Bosonic 
$U^{a}{}^{}_{b}$-integrations, which are compact directions. Boundary terms can
only arise from Bosonic $C^{a}{}^{}_{b}$-integrations.) Perhaps surprisingly, 
the pertinent Fermionic ``$\brst$'' transformation \e{delta2} needed for the
localization argument is {\em not} the BRST operator. {}Furthermore, it turns out that ``$\brst$'' is {\em not} nilpotent. (One of our initial motivations was
to investigate whether ``$\brst$'' and the BRST operator would coincide, or 
not, and whether ``$\brst$'' would be nilpotent, or not.)
In general, the non-nilpotency implies, among other things, that an
$\brst$-exact quantity is {\em not} necessarily $\brst$-closed, and that a
product of $\brst$-exact quantities is {\em not} necessarily $\brst$-exact nor
$\brst$-closed. Nevertheless, the square $\brst^{2}$ is still a linear
derivation, $\brst^{2}(f g)\!=\!(\brst^{2}f) g \!+\! f (\brst^{2}g)$, with
\bea
\brst^{2}U = U(C^{2}\!+\!H^{\prime}) , && 
\brst^{2}H = [H, C^{2}\!+\!H^{\prime}] , \qquad
\brst^{2}C = [H, C]^{\prime} , \label{delta3a} \cr
\brst^{2}U^{\dag} = -(C^{2}\!+\!H^{\prime})U^{\dag} ,  &&  
\brst^{2}\Pi = [\Pi, C^{2}\!+\!H^{\prime}] . \label{delta3b}
\eea 
Therefore we will restrict ourselves to consider the subalgebra of integrands 
$f\!=\!f(U,C)$ with $\brst^{2}f\!=\!0$. In particular, we will consider a 
Fermionic function $\psi\!=\!\psi(U,C)$ given by 
\beq
\psi := \tr(H C) = \tr(H^{\prime} C) , \qquad
\brst\psi = \tr\left(H^{\prime}\brst C - C \brst H\right)
 = \tr\left(H^{\prime}H^{\prime}-C [H, C]\right) , \label{psiexplicit}
\eeq
\beq
\brst^{2}\psi 
= \tr\left(H^{\prime}\brst^{2}C + C \brst^{2} H\right)
= \tr\left(H^{\prime}[H, C] + C[H, C^{2}\!+\!H^{\prime}]\right) = 0 .
\eeq
The above cohomological consideration shows that the partition function
\beq
{\cal Z}^{(t)} = \int^{}_{U\in{\rm End}(V)} \!\!\!\!\!\!\!\!
[dU][dU^{\dag}][d\Pi][dC]e^{\ih S^{(t)}}  \label{izintt} 
\eeq
with action
\beq
S^{(t)} = S^{}_{1} - \frac{1}{2t^{2}}\brst\psi
=\tr\left(HB - \frac{1}{2t^{2}}H^{\prime}H^{\prime} + \Hf C[B+t^{-2}H,C] 
+ (U^{\dag}U\!-\!{\bf 1})\Pi\right)   \label{esst}
\eeq
cannot depend on the parameter $t$, because $\brst S^{}_{1}\!=\!0$ and 
$\brst^{2}\psi\!=\!0$. In the limit $t\to\infty$, the partition function 
$\lim^{}_{t\to\infty} {\cal Z}^{(t)} = {\cal Z}^{}_{1}$ is just the
sought--for integral \e{izint1}. ({}For each Bosonic Gaussian
$C^{a}{}^{}_{b}$-integration, one might have to adjust the 
$C^{a}{}^{}_{b}$-integration contour as a function of $t$ and $U$ to ensure
that the $C^{a}{}^{}_{b}$-integral remains exponentially damped in the integral
\e{izintt}. The value of the Gaussian $C^{a}{}^{}_{b}$-integral is unchanged 
under such shift of $C^{a}{}^{}_{b}$-integration contour.)  

\noi
Often in the literature, one only provides an implicit existence argument that
a pertinent Fermion $\psi$ with $\brst^{2}\psi\!=\!0$ exists. Here we actually
have an explicit formula \e{psiexplicit} for $\psi$.

\subsection{Localization}

\noi
Let us scale the off--diagonal Fermionic integration variables $C \to t C$ 
with the number $t$. This produces a Jacobian factor
\beq
\prod_{1\leq a\neq b\leq N} t^{\left[(-1)^{\eps(C^{a}{}^{}_{b})}\right]}
=\prod_{1\leq a\neq b\leq N} t^{\left[-(-1)^{\eps_{a}+\eps_{b}}\right]}
= t^{2N^{}_{0}N^{}_{1}-N^{}_{0}(N^{}_{0}-1)-N^{}_{1}(N^{}_{1}-1)} .
\label{tjac}
\eeq
Recall that for an arbitrary complex supernumber $z$ of Grassmann--parity
$\eps(z)$, one has 
\beq
\lim_{\pm t\to 0^{+}} t^{-2\left[(-1)^{\eps(z)}\right]}
\exp\left[-\frac{z\bar{z}}{2t^{2}}\right]
=\left\{\begin{array}{ccc}2\pi\delta^{2}(z) &\for&\eps(z)\!=\!0  
\cr -\Hf z\bar{z} &\for&\eps(z)\!=\!1 \end{array}\right\}
\sim \delta^{2}(z) , \label{deltafct}
\eeq
where we have suppress an overall numerical factor in the last expression of 
\eq{deltafct}. (In detail, the limit notation $\pm t\to 0^{+}$ in \eq{deltafct}
is supposed to mean that the limit should be performed in such a way that 
$|Im(t)|\!<\!|Re(t)|$ for Bosonic $z$.)
One now let $H^{\prime a}{}^{}_{b}$ in the $S^{(t)}$ action \e{esst} play the 
r\^ole of $z$. (This is possible since $H^{\prime}$ is Hermitian, as a result 
of the matrix $A$ being Hermitian.)
Adapting \eq{deltafct} to the oscillatory ${\cal Z}^{(t)}$ integral \e{izintt},
one is interested in the limit $\pm e^{-\frac{i\pi}{4}}t\to 0^{+}$.
\beq
\lim^{}_{\pm e^{-\frac{i\pi}{4}}t\to 0^{+}} {\cal Z}^{(t)} 
\sim \int^{}_{U\in{\rm End}(V)} \!\!\!\!\!\!\!\!
[dU][dU^{\dag}][dC]e^{\ih \tr\left(HB + \Hf C[H,C]\right)} 
\delta(H^{\prime})\delta(U^{\dag}U\!-\!{\bf 1}) ,
\label{izint2}
\eeq
Equation \e{izint2} shows that the integral localizes on the constraint 
$H^{\prime}\approx 0$, which, in turn, is just the stationary surface 
\e{uclassol}!
Hence, in order to evaluate the  integral \e{izint2},
it is enough to consider an infinitesimally small tubular $U$-neighborhood of 
the stationary surface \e{uclassol}. One must sum over all possible
instanton sectors labelled by $\sigma\in S^{}_{N^{}_{0}}\!\times\!S^{}_{N^{}_{1}}$.
{}For a given permutation $\sigma\in S^{}_{N^{}_{0}}\!\times\!S^{}_{N^{}_{1}}$,
one may hence parametrize the $U$-variable as 
\beq
 (\Xi,K) \longrightarrow  U = P^{}_{\sigma}e^{i\Xi}e^{K}
= e^{iP^{}_{\sigma}\Xi P^{}_{\sigma}}P^{}_{\sigma}e^{K}
  , \label{changeofvar}
\eeq
where $\Xi$ is a real, diagonal matrix, and where $K\!\in\!\cA$ is an 
off--diagonal Bosonic matrix, which may be taken to be infinitesimally small. 
One calculates 
\beq
H = U^{-1}AU = e^{-K} P^{}_{\sigma}AP^{}_{\sigma}e^{K}
= P^{}_{\sigma}AP^{}_{\sigma}+ \left[P^{}_{\sigma}AP^{}_{\sigma},K\right] 
+ {\cal O}\left( K^{2}\right) .
\eeq
The Cartan torus $e^{i\Xi}$ in 
\eqs{uclassol}{changeofvar} just reflects a compact $U(1)^{N}$ gauge symmetry. 
The integration over the diagonal/gauge directions $\Xi$ therefore yields 
the volume $\vol\left(U(1)^{N}\right)\!=\!(2\pi)^{N}$ of the Cartan torus 
$U(1)^{N}$, which we ignore, since we are not interested in overall numerical 
factors. 
The integral \e{izint2} becomes
\bea
\lim^{}_{\pm e^{-\frac{i\pi}{4}}t\to 0^{+}} {\cal Z}^{(t)}  
&\sim& \sum_{\sigma\in S^{}_{N^{}_{0}}\!\times\!S^{}_{N^{}_{1}}}
\int \! [dK][dK^{\dag}][dC] 
e^{\ih \tr\left(P^{}_{\sigma}AP^{}_{\sigma}B 
+ \Hf C\left[P^{}_{\sigma}AP^{}_{\sigma},C\right]\right)} 
\delta([P^{}_{\sigma}AP^{}_{\sigma},K])\delta(K^{\dag}\!+\!K) \cr
&=& \sum_{\sigma\in S^{}_{N^{}_{0}}\!\times\!S^{}_{N^{}_{1}}}
\int \!dK[dC]  e^{\ih \tr\left(P^{}_{\sigma}AP^{}_{\sigma}B 
+ \Hf C\left[P^{}_{\sigma}AP^{}_{\sigma},C\right]\right)} 
\frac{\delta(K)}
{\Delta^{2}_{}(\lambda^{}_{\sigma(a)})} \cr
&\sim& \sum_{\sigma\in S^{}_{N^{}_{0}}\!\times\!S^{}_{N^{}_{1}}} 
\frac{e^{\ih (-1)^{\eps_{a}}\lambda^{}_{\sigma(a)}\mu^{}_{a}}}
{\Delta(\lambda^{}_{\sigma(a)})}
=\sum_{\sigma\in S^{}_{N^{}_{0}}\!\times\!S^{}_{N^{}_{1}}}(-1)^{\sigma}
\frac{e^{\ih (-1)^{\eps_{a}}\lambda^{}_{\sigma(a)}\mu^{}_{a}}}
{\Delta(\lambda^{}_{a})} \cr
&=&\frac{
\det\left(e^{\ih\lambda^{}_{a}\mu^{}_{b}}\right)^{}_{1\leq a,b\leq N^{}_{0}} 
\det\left(e^{-\ih\lambda^{}_{a}\mu^{}_{b}}\right)^{}_{N^{}_{0}+1\leq a,b\leq N}}
{\Delta\left(\lambda^{}_{a}\right)} , \label{superhciz}
\eea
in agreement with the superversion of the Harish--Chandra--Itzykson--Zuber 
formula \cite{Alfaro:1994ca,Guhr:1996mx} up to a numerical factor.

\end{document}